THE PATENTIST
LIVING
LITERATURE REVIEW

# Why do firms patent?

Gaétan de Rassenfosse

Holder of the Chair of Science, Technology, and Innovation Policy
École polytechnique fédérale de Lausanne, Switzerland.

**Purpose**
This article is part of a Living Literature Review exploring topics related to intellectual property, focusing on insights from the economic literature. Our aim is to provide a clear and non-technical introduction to patent rights, making them accessible to graduate students, legal scholars and practitioners, policymakers, and anyone curious about the subject.

**Funding**
This project is made possible through a Living Literature Review grant generously provided by Open Philanthropy. Open Philanthropy does not exert editorial control over this work, and the views expressed here do not necessarily reflect those of Open Philanthropy.

## Why do firms patent?

Gaétan de Rassenfosse
École polytechnique fédérale de Lausanne, Switzerland.

A previous article in this series highlighted that not all patentable inventions are patented, prompting the question of why firms choose to patent or not. The present article explores the factors that affect the propensity to patent. Examining these factors helps clarify the conditions under which the patent system operates effectively.

### Is patent protection worth it?

Every firm considering patenting an invention asks itself this question, seeking to assess the costs and benefits of patent protection. The costs are generally known, incurred upfront, and encompassing both monetary and non-monetary factors. In contrast, the benefits are more challenging to identify and quantify and remain uncertain. Moreover, not all patent applications are approved, meaning that firms could incur the costs of patent protection without realizing the benefits if their application is rejected. Let us analyze the cost-benefit tradeoff of patent protection.

On the cost side, the monetary expenses are straightforward to quantify. They encompass the patent attorney fees and the administrative charges, with total costs in the United States typically ranging between $5,000 and $30,000—although there is obviously no strict maximum limit. However, because patent rights are jurisdictional, having legal effect only in the jurisdiction that issues them, firms seeking international patent protection must file patents—and bear the associated costs—in every country where they seek protection. These costs can escalate quickly and be prohibitively expensive for small and medium enterprises.

The non-monetary cost primarily relates to the disclosure requirements. To secure a patent, an inventor must disclose the workings of the invention, which become public information. This transparency may assist competitors in designing around the patent or advancing alternative technologies. Furthermore, once the patent expires, competitors can copy the disclosed invention. The disclosure requirement is one strong reason why some firms opt for secrecy instead of patenting. Stories of inventions kept secret abound, although examples commonly found primarily relate to the food and beverage industry, including the Coca-Cola formula, KFC Original Recipe of '11 herbs and spices,' and the Pastéis de Belém. However, survey data consistently indicate that a larger share of companies valued secrecy over patents across many industries—a tendency especially pronounced among smaller firms (Arundel 2001, Brouwer and Kleinknecht 1999).

### What are the benefits of patent protection?

A patent grants its holder the right to exclude others from making, using, selling, or importing the patented invention. Patents help firms 'appropriate the returns' or 'capture the value' of their innovations—terms coined by economists and management scholars, respectively, to indicate that patents shield inventions from imitation. However, as we will discuss, there are many ways to leverage patents, resulting in diverse motives for filing—and a broad range of associated benefits. To explore these benefits, researchers have surveyed patent applicants about their reasons for filing. These studies cover, among others, established U.S. firms (Cohen et al. 2001), U.S. high-tech startups (Graham et al. 2009), European small and medium enterprises (de Rassenfosse 2012), and firms in Switzerland (Harabi 1995), France (Arundel and Kabla 1998), Germany (Blind et al. 2006) and Spain (Martinez and Penas 2013).

The surveys reveal a variety of overlapping and sometimes context-specific reasons why firms seek patents. Alongside the traditional goal of protecting their inventions, companies also file patents strategically to block competitors, restricting access to patented content even when they have no intention of using it themselves. An 'offensive' patent blocking strategy might involve disrupting rivals' R&D or product pipelines, while a 'defensive' blocking approach focuses on preserving a firm's freedom to operate and avoiding infringement lawsuits. Tactics such as building 'patent fences' or 'patent walls,' or contributing to 'patent thickets,' support these objectives.

Another key motive for filing patents is to use them as negotiation tools or 'bargaining chips' in inter-company agreements and collaborations, such as cross-licensing deals, joint ventures, and standard-setting activities. Holding patents can significantly strengthen a firm's bargaining position and help deter litigation, with such strategy sometimes leading to what is known as a 'patent arms race' (Chien, 2010).

Despite their variety, these motives for filing patents rest on a fundamental aspect: the exclusion right granted by patents. This legal power underpins every strategy discussed above—from protecting inventions and blocking competitors to facilitating negotiations—by ensuring that the patent holder can control how the invention is used. However, there is also a range of patent uses that do not rely on the exclusion right, as discussed below.

**Benefits extend beyond the exclusion rights**

Firms might file patent applications solely to create prior art. This approach stops others from patenting the same invention and ensures that the firm can continue to use its own technology freely. In these cases, the objective is to establish a documented record that limits future patent claims, rather than to secure patent protection.

Patents can also serve as signals of a firm's quality and innovation (Long 2002). A well-known example is Audi's 2006 A6 advertisement, which compared the number of patents held by NASA with those held by Audi to highlight the car's advanced technology. The commercial ended with the line "To date, Nasa have filed 6,509 patents. To get to the A6, Audi have filed 9,621 patents." Although Audi likely did not pursue patents specifically for the ad, the campaign illustrates that firms can use patents to enhance their reputation.

Firms, especially smaller ones and startups, may also file patents to bolster their attractiveness to investors, but patents serve two distinct roles in this regard. On one hand, the exclusion right directly protects a firm's market position and enhances its chances of survival, something that investors value. On the other hand, the mere presence of patents can act as a signal to investors that a company is innovative and professionally managed, lending credibility to the firm's technology novelty. This 'signal' increases investor confidence and operates irrespective of the way the exclusion right will be used.

Finally, patents are sometimes used as performance indicators for researchers and R&D teams, offering a measure of innovation and productivity. Furthermore, being named as an inventor on a patent not only can lead to additional compensation and career advancement but may also reinforce an employee's value within the company. Some organizations showcase their patents in the form of 'patent walls,' underscoring their commitment to innovation and celebrating the creative contributions of their personnel.

Overall, protecting innovations from imitation and blocking competitors consistently emerge as key drivers across different firm types and countries. However, strategic uses such as negotiation and licensing, signaling value, and securing financing are also important motivations, with their significance varying depending on the specific characteristics of the firms and their environments.

**The challenge of evaluating the benefits of patents**

Although there may be numerous potential benefits, quantifying their value at the time of the patenting decision is highly speculative. This uncertainty complicates the choice of whether to patent an invention. First, there is no guarantee that the application will be granted. If it is rejected, the applicant incurs all costs—including the cost of disclosure, as the application is published—without gaining any of the anticipated benefits. Moreover, while the grant decision is based on an invention's technical merits, its actual economic payoff hinges on market potential, creating a disconnect between the likelihood of grant and the value ultimately realized. Finally, because the rewards of patent protection materialize in the future, firms must predict how technology and competition will evolve, adding yet another layer of uncertainty.

Considerations about the 'option value' of patents further complicate the patenting decision. Even if a simple cost-benefit analysis may not justify the immediate expense of filing a patent, the patent's option value can still make it a worthwhile investment. Once a patent is filed, the firm secures the right to benefit from its protection, much like holding a call option on future innovations. By contrast, choosing not to file means permanently relinquishing this option, even if market conditions later become more favorable for enforcing or monetizing the technology.

Given this uncertainty, patents are often likened to 'lottery tickets' that offer potentially significant, albeit speculative, rewards to their holders. Moreover, the ultimate value of a patent frequently depends on its ability to stand up in court, meaning that its benefits are sometimes conditional on successful enforcement. Several studies suggest that a considerable number of U.S. patents would be deemed 'invalid' if challenged in court (a topic we will explore in a separate article), further adding to the hypothetical nature of the rewards associated with patent protection.

**Barriers to patenting**

In addition to examining why firms pursue patents, scholars have investigated factors that discourage patenting. Unsurprisingly, high costs and the mandatory disclosure of key information stand out as major barriers. Beyond these concerns, research also points to the limited effectiveness of patents in preventing imitation, given that competitors can often 'invent around' existing patents or that enforcement can prove both difficult and costly.

Another set of barriers involves the complexity and time of the patenting process. Because it can be tedious, firms may be discouraged from pursuing patents altogether. Moreover, the delay between filing and issuance—often spanning several years—can be particularly problematic in rapidly evolving fields, where a patent might become obsolete before it is granted.

Lastly, some innovative firms, especially those with limited experience or resources, may not realize that their inventions are patentable in the first place. This knowledge gap can arise from inadequate internal processes for identifying intellectual property or from a general lack of familiarity with patenting strategies and best practices. As a result, potentially valuable innovations might go unprotected simply because firms remain unaware of their patentability or are unsure of how to navigate the system.

**If firms don't patent, how do they capture the value of their inventions?**

While patents offer a recognized path to securing returns on an invention, firms often rely on a variety of other strategies that may be more suitable for specific industries, innovation types, or organizational circumstances. One obvious alternative is secrecy, as previously discussed, which works especially well for inventions shielded from public view, such as manufacturing processes. Moreover, if a product or process is inherently complex, firms can rely on that complexity to discourage imitation. Inventions that are costly and time-consuming to imitate form a natural barrier that deters potential competitors.

Another strategy is to gain a lead time advantage by introducing an innovation before competitors, allowing the firm to build market presence and cultivate brand loyalty. This approach is particularly valuable for smaller companies, which often emphasize speed to market as a core protection mechanism. Being first to market also enables a firm to refine production processes ahead of potential imitators, helping it move quickly down the learning curve, thereby reducing costs and boosting efficiency—advantages that can be challenging for competitors to replicate. Likewise, by regularly upgrading or improving its offerings, a firm can stay one step ahead, rendering existing products or processes obsolete before imitators can catch up.

Innovators can also capture value by offering complementary products or services, drawing on capabilities in manufacturing, marketing, sales, and service. Even if the core innovation is easily replicated, a firm's established network, specialized expertise, and infrastructure can still provide a crucial competitive edge.

It is important to recognize that these strategies are not mutually exclusive and can often be combined for enhanced effect. For example, a strong patent portfolio can work in tandem with efforts to build brand loyalty, while secrecy and formal patent protection may coexist even within the same product line. The protection of a novel beauty cream might rely on patents for individual compounds and secrecy for the specific mixture.

To conclude, even amid questions of costs, validity and effectiveness, patents continue to offer strategic advantages that many firms find indispensable. Firms patent not only to deter imitators but also to create new opportunities—through collaboration, financing, and reputation-building—underscoring why patents remain central to modern innovation strategies.